# Semantic Intelligence Against CSAM: The PreventCSA@EU Ontology Framework for Classification and Investigation

*Elias Tzortzakakis, Emmanouela Kokolaki[1], Evangelia Daskalaki and Paraskevi Fragopoulou*

Foundation for Research and Technology - Hellas (FORTH), Institute of Computer Science
N. Plastira 100, Vassilika Vouton, GR-70013 Heraklion, Crete, Greece
{tzortzak, kokolaem, eva, fragopou}@ics.forth.gr

## Abstract

This work presents the PreventCSA@EU ontology, a semantically grounded framework designed to support the identification, classification, annotation, and analysis of online Child Sexual Abuse and Child Sexual Exploitation Material (CSAM/CSEM). The growing circulation and dissemination of CSAM/CSEM across digital environments, combined with inconsistencies in legal definitions and classification practices across jurisdictions, highlights the need for semantically interoperable frameworks capable of supporting cross-organizational cooperation and automated processing. The proposed ontology is developed through a systematic review and comparative analysis of existing CSA/CSE-related, metadata-oriented, and investigative ontologies and taxonomies, with its primary design aimed at addressing the operational needs and domain-specific requirements of national LEA Directorates. It introduces a hierarchical semantic model built around core entities such as Media Object, Content, Person, Depiction, and Investigative Report, while enabling structured alignment with INHOPE UCS labels, Dublin Core–DCMI Metadata Terms, and Schema.org. The proposed framework emphasizes ontology-driven interoperability for structured annotation and analysis of CSA/CSE-related data, supporting consistent classification, child identification, and investigative processes for offender prosecution. The design aims extend existing classification approaches with additional conceptual structures for database conceptualization, process modeling, and ontology-driven data management. By integrating established classification standards with a novel hierarchical ontology, the proposed framework enhances cross-system compatibility, with particular relevance to emerging EU-level data infrastructures, including the envisaged EU Centre database under the proposed Child Sexual Abuse Regulation (CSAR).

## 1. Introduction

The effective identification, classification, and analysis of online Child Sexual Abuse (CSA) and Child Sexual Exploitation (CSE) material constitute a critical challenge for law enforcement agencies, non-governmental organizations, hotlines, and technology industry stakeholders worldwide. The increasing volume and heterogeneity of digital content, combined with significant variations in national legal definitions and classification practices, have highlighted the need for semantically grounded frameworks capable of supporting cross-jurisdictional cooperation and automated processing. In this context, ontologies have emerged as a powerful means for formally representing domain knowledge, enabling semantic interoperability and consistent conceptualization across systems.

This paper presents the PreventCSA@EU ontology, a formal specification of the concepts, entities and relationships within the domain of CSA and CSE [1]. It provides an explicit conceptualization of the domain, supporting semantic representation and knowledge sharing. The proposed ontology [2] for CSA and CSE serves as a structured and semantically precise framework for organizing, categorizing and interrelating the concepts, entities and relationships relevant to online CSAM and CSEM.

[1] Corresponding author: kokolaem@ics.forth.gr; ORCID 0009-0003-9982-7699;

The ontology was developed within the framework of the European project "Towards a Coordinated and Cooperative Effort for the Prevention of Child Sexual Abuse at a European level", which aims to strengthen cooperation among national authorities, law enforcement agencies, online service providers and EU institutions in preventing and combating child sexual abuse, online child sexual exploitation and grooming. A central objective of the project is the development of an Annotated Hash Database (AHDB) for child sexual abuse material (CSAM/CSEM), intended to support the detection, classification, reporting and removal of harmful content through a standardized approach. Within this context, the PreventCSA@EU ontology was designed to provide the semantic foundation for the database, ensuring consistency and interoperability across CSAM/CSEM management workflows.

In addition, this paper provides a comprehensive state-of-the-art review, alongside a systematic comparison and classification of existing ontologies and related frameworks within the CSA/CSE domain. By organizing these approaches into distinct categories based on their objectives and functional orientation, the analysis highlights their respective strengths, limitations, and areas of application. This comparative perspective not only situates the PreventCSA@EU ontology within the broader research and operational landscape, but also clarifies its contribution in addressing existing gaps and advancing semantic interoperability in this domain.

Following the presentation of the PreventCSA@EU ontology, its alignment with the core elements of the Universal Classification Schema (UCS), introduced by INHOPE within the Global Standard Project, is illustrated. This design choice aims to ensure semantic interoperability, while extending the ontology with additional conceptual structures required for database conceptualization, process modeling, and ontology-driven data management. By combining state-of-the-art classification standards with an entirely novel organizational and hierarchical structure, the proposed ontology supports extensibility, ultimately aiming to ensure compatibility with the envisaged EU Centre database [3], as outlined in the Proposal for a Child Sexual Abuse Regulation (CSAR).

From a broader perspective, this work showcases state-of-the-art developments in ontology-based modeling, demonstrating advancements specific to CSA/CSE, while also illustrating general principles and best practices in ontology design. It highlights the role of a domain ontology as a unifying conceptual layer that bridges international classification standards and operational database systems. Building upon the reviewed state of the art, the PreventCSA@EU ontology exemplifies an approach in which semantic clarity, alignment with established standards and enhanced reasoning capabilities foster consistency, interoperability and cross-organizational collaboration in a highly sensitive domain.

## 2. State-of-the-Art

The landscape of ontology development within the domain of Child Sexual Abuse (CSA) and Child Sexual Exploitation (CSE) is both complex and multifaceted, comprising a variety of ontologies that address different objectives. Existing research can generally be grouped into three main categories: CSA/CSE-focused ontologies, metadata-oriented ontologies, and investigative ontologies. The primary goal of CSA/CSE focused ontologies is to facilitate better victim identification and more effective processing of CSAM by law enforcement officers, hotline analysts, and technology industry professionals. Metadata-related ontologies provide a formal structure for defining, categorizing, and interrelating metadata elements, which describe the attributes and context of data, digital content, or resources. Investigative ontologies provide a standardized, structured approach for capturing and exchanging the information commonly examined by humans and systems during investigations involving digital evidence.

### 2.1 CSA/CSE Related Ontologies

CSA/CSE related ontologies are designed to strengthen victim identification efforts and optimize the processing of CSAM by key stakeholders, including hotline analysts, law enforcement, and the tech industry. By providing standardized classification systems, these ontologies support the development of tools and frameworks for identifying, analyzing, and responding to abusive activities online. Below are several schemas that have served as foundational drivers in the development of the Pre-ventCSA@EU Ontology.

***Universal Classification Schema***

INHOPE launched in 2023 a common schema for the classification of child sexual abuse and exploitation material [4] . This schema was created as a part of the Global Standard Project and aims to create a shared language for identifying and categorizing content across industry, non-governmental organizations (NGOs) and law enforcement. By addressing variations in the legal definition of CSAM across jurisdictions, it enables smoother data exchange and more effective coordination. In addition, its unified structure enables the effective deployment of machine learning technologies for CSAM detection and supports the creation of annotated datasets for training automated CSAM detectors, thereby enhancing victim identification and strengthening investigative processes.

The INHOPE Classification Schema v3.0 consists of three main classes, namely the *Categorization Elements*, the *Investigative Elements* and the *Demographic Elements*. The Categorization Elements class consists of label classes and labels that focus primarily on what is exclusively contained within the media. The Investigative Elements class includes label classes and labels that capture contextual characteristics and additional information that do not inherently affect the material's legality but are valuable for investigative purposes, such as victim identification and reporting. The Demographic Elements class includes classes that describe key characteristics of the individuals depicted in the media content.

Due to usage restrictions, the underlying Universal Classification Schema (UCS) is not publicly available, and its detailed content cannot be disclosed in this paper. Nevertheless, the manuscript presents the conceptual framework and methodology in sufficient detail to support understanding of the proposed approach, while acknowledging that these restrictions may limit the extent of independent verification. Access to the UCS may be granted upon approval of a formal request submitted to INHOPE. INHOPE also supports the implementation of the UCS by providing integration guidance, assistance in tailoring the schema to reporting workflows, and training sessions.

***Terminology Guidelines for the Protection of Children from Sexual Exploitation and Abuse (Luxembourg Guidelines)***

In September 2014, an Interagency Working Group (IWG) comprising representatives from key stakeholders, including law enforcement and international bodies, was established. Leveraging the expertise of IWG representatives and their respective organizations, a comprehensive examination and analysis on terminology and definitions was conducted over the course of more than a year. The interagency effort culminated in the publication of the "Terminology Guidelines" in 2016 [5]. The document compiles a set of terms related to the protection of children from sexual exploitation and sexual abuse, reflecting the language commonly employed by professionals and international agencies in efforts to prevent and address such violations. Its primary purpose is to provide a standardized framework that ensures precise and consistent terminology across advocacy, policy and operational contexts globally.

The initiative emerged in response to longstanding inconsistencies and conceptual ambiguities in the terminology used to describe sexual exploitation and sexual abuse of children. Despite the existence of certain definitions in international and regional legal instruments, divergent interpretations and inconsistent applications of key terms have persisted across jurisdictions and institutions. Such discrepancies have not only complicated legislative drafting and policy development, but have also hindered effective data collection, comparative research, and the measurement of impact. In cross-border contexts, these challenges are further amplified, as the absence of shared conceptual understandings undermines coordinated international responses.

A central objective of the Terminology Guidelines was therefore to enhance conceptual clarity and promote harmonized usage of terms across languages and regions. The document systematically explains the linguistic meaning and practical implications of each term, identifies instances in which caution is warranted, and discourages terminology that may perpetuate harmful stereotypes or weaken child rights protection. In doing so, it draws upon definitions contained in international and regional treaties, as well as interpretative guidance from human rights monitoring bodies and relevant resolutions and recommendations issued by international organizations.

The Guidelines further articulate principled criteria for the inclusion of terms, prioritizing those that possess a recognized legal basis, are widely operationalized in professional practice, or generate significant misunderstanding concerning children's rights and entitlements under international law. Structurally, the document progresses from foundational legal concepts (e.g. the definition of the "child" and overarching notions of sexual violence, sexual abuse, and sexual exploitation) to more specific manifestations (including child prostitution, child pornography, and exploitation occurring in the contexts of travel, tourism, and digital environments). This systematic progression reflects an effort to anchor specific phenomena within a coherent conceptual and normative architecture.

Importantly, the Guidelines explicitly acknowledge the evolving nature of sexual exploitation and sexual abuse, particularly in light of rapid developments in information and communication technologies (ICTs). Emerging practices, such as online grooming and live-streamed sexual abuse, highlight gaps in existing definitions. By providing preliminary analysis of these phenomena and stressing the need for regular review, the Guidelines frame terminology as a dynamic normative tool, ensuring that legal and policy responses remain coherent, responsive, and aligned with contemporary threats to the sexual integrity of children.

***Towards a Global Indicator of Unidentified Victims in Child Sexual Exploitation Material***

The report "Towards a Global Indicator on Unidentified Victims in Child Sexual Exploitation Material" [6] presents findings from a two-part analysis of the multi-country dataset included in the International Child Sexual Exploitation (ICSE) Database, alongside consultations with law enforcement personnel regarding the identification of victims and offenders depicted in Child Sexual Abuse Material (CSAM) and Child Sexual Exploitation Material (CSEM).

This Database is maintained at INTERPOL and forms part of a larger program of the ICSE Database enhancement activities, financed by the European Union and carried out between 2016 and 2018 under the title International Child Sexual Exploitation (ICSE) Database Connectivity and Awareness Raising Enhancements (I-CARE) Project, with two partners namely INTERPOL and ECPAT.

The report provides insights into the profiles of unidentified child victims and their abusers, based in visual examination of 800 series of videos and images which were randomly selected for a more detailed analysis, including age, gender and the severity of sexual victimization, and further present the results of case related metadata for both identified and unidentified cases recorded in the ICSE Database. The study ultimately developed a descriptive profile of unidentified children depicted in CSAM/CSEM in the ICSE and shed light on the functionality and the content of the ICSE Database. The report is based on the Luxembourg Guidelines analyzed above, but it also introduces other terminologies that are widely used in the ICSE Database of Interpol.

Among the various elements that comprise the ICSE Database, the most important addition is the inclusion of the term "Baseline". Given that the definitions of CSAM vary across different jurisdictions, "Baseline" serves as a CSAM classification and aims to establish an international standard that identifies CSAM which is illegal worldwide. The ultimate goal is to provide industry stakeholders with a hash list of such media, enabling them to detect its presence on their systems, remove it, and report it to the competent authorities. Other ICSE elements include series, ungrouped media, metadata, and investigation (**Table *1*** below).

| ICSE Term | Notes |
|---|---|
| Series | A series comprises images and/or videos that are meaningfully associated from an investigative perspective, for example because they depict the same victim or originate from the same crime scene. Grouping all related media within a single series is a fundamental step in the victim identification process. A series may also contain only a single image when that image provides sufficient information to support victim identification or when it depicts a child who has already been identified. |
| Ungrouped Media | Media stored in the database that have not been assigned to any series. |
| Metadata | Descriptive information associated with a media item and/or series that is entered by an ICSE user during the upload process. Such information may include the national case reference, information describing where and how the media was discovered, the location and duration of the abuse, and the age of the victim(s) at the time of the abuse. |
| Investigation | An investigation is created in the ICSE Database by a user to notify INTERPOL and all participating countries that a child and/or offender has either been identified within their jurisdiction or is currently the subject of an active investigation. A single investigation may be associated with multiple series. |
| Baseline | The Baseline category was established within the ICSE Database as an international standard intended to identify the most severe forms of Child Sexual Abuse Material (CSAM). Media classified as Baseline are expected to be considered illegal in all jurisdictions where legislation addressing CSAM is in force. The purpose of this classification is to enable the sharing of corresponding hash values with industry partners, allowing them to detect such material within their networks or systems and to facilitate its reporting and removal.<br><br>For media to qualify as Baseline, it must satisfy all of the following criteria:<br>It depicts a real child rather than an artificially generated image.<br>The child is prepubescent, exhibiting no or only the earliest signs of puberty and appearing to be younger than 12–13 years of age.<br>The child is involved in, or witnessing, sexual activity, or the material places a clear visual emphasis on the child's genital or anal area. |

**Table 1**. Elements of the ICSE Database. Adapted from [6].

***The Tech Coalition's Industry Classification System***

The Tech Coalition is an alliance of tech companies dedicated to combating child sexual exploitation and abuse online. To support its objective, its members have developed and adopted a voluntary open-source image classification system [7] which is widely used by electronic service providers to categorize images and videos depicting apparent child sexual abuse and exploitation. This classification framework is commonly applied when providers submit CSAM/CSEM to the National Center for Missing and Exploited Children (NCMEC). The use of a standardized classification approach facilitates the efficient prioritization and review of the large volume of reported material by NCMEC and other authorized organizations, helping to identify the most severe cases more effectively.

The classification system works like a “matrix”, with one axis representing the apparent age of the victim. Images and videos of prepubescent children are marked as “A”, while content depicting victims appearing to be under 18, but are not postpubescent, is marked as “B”. The second axis categorizes the conduct depicted in the content: images or videos that include a sexual act are designated “1”, while the content that qualifies as “child pornography” under the federal law, but does not depict a sex act (instead depicting a “lascivious exhibition”) is classified as “2”.

### 2.2 Metadata Related Ontologies

Metadata related ontologies provide a structured framework for organizing and connecting metadata elements that describe the properties and contextual information of data and resources. These ontologies support interoperable representations across different systems, enhancing data reusability.

***Dublin Core – DCMI Metadata Terms***

The Dublin Core Metadata Initiative is an organization that supports innovation in metadata design. One of its key contributions is the management and development of specifications and metadata terms namespaces. The authoritative DCMI Metadata Terms specification is one of these namespaces, which includes the fifteen core terms of the Dublin Core™ Metadata Element Set, along with numerous additional properties, classes, datatypes, and vocabulary encoding schemes.

DCMI metadata terms are expressed in RDF vocabularies for Linked Data applications. However, the same terms can also be used in non-RDF environments, including XML, JSON, UML and relational databases, without relying on the global identifiers or formal RDF semantics. In such cases, users can treat the domain, range, subproperty, and subclass relationships as guidelines for usage, focusing on the natural language text of the definitions, usage notes, and examples.

Each term is assigned a Uniform Resource Identifier (URI). Term URIs point to the DCMI Metadata Terms, while programmatic references via RDF applications lead to one of four associated RDF schemas. Each RDF’s scope corresponds to a DCMI namespace or set of DCMI metadata terms identified using a common base URI, as specified in the DCMI Namespaces Policy. The URIs for DCMI namespaces are frequently declared as prefixes to simplify queries and schemas. In addition, DCMI metadata terms are provided in machine readable formats such as RDF/XML and Turtle, allowing their use in Semantic Web and Linked Data applications. Since 2000, Dublin Core terms have been assigned Persistent URLs (PURLs), which ensure that references to the terms remain stable over time. These URIs allow RDF-based applications to directly integrate the terms, while non-RDF users can still use the URIs for consistent identification and referencing.

Moreover, DCMI maintains separate namespaces for different sets of terms, such as the DCMI Type Vocabulary for classifying resources, the DCMI Metadata Terms for general descriptive metadata, and additional vocabularies for encoding schemes and controlled terms. This modular structure allows users to select and combine terms as needed, while preserving semantic consistency and interoperability across datasets and applications.

***Schema.org***

Schema.org is a collaborative initiative [8] that provides a shared vocabulary for describing structured data on the Web. Its vocabulary can be used with many different encodings, including RDFa, Microdata and JSON-LD. The vocabulary defines entities together with the relationships between them and has been designed to be extensible through a well-defined extension mechanism. Major search engines and web platforms, includ-ing Google, Microsoft, Pinterest, and Yandex, make extensive use of these vocabularies to support the integration and exchange of structured data. The main schema.org hierarchy consists of a collection of types (classes), each associated with one or more parent types, forming a structured classification system.

The initiative was launched in 2011 by the major search engines Google, Bing, and Yahoo, and was later joined by Yandex, with the aim of providing a single, integrated schema covering a wide range of topics, such as people, places, events, products, and offers. By standardizing the vocabulary, webmasters can annotate content once while benefiting multiple consumers across different platforms. Since its inception, Schema.org has expanded from 297 classes and 187 relations to over 600 classes and nearly 1,000 relations, organized in a hierarchical structure with polymorphic relations that allow flexibility in describing complex and overlapping domains.

A central principle of Schema.org's design is pragmatism, balancing ease of use for webmasters with machine readability. Multiple syntaxes are supported, including RDFa, Microdata, and JSON-LD, to accommodate diverse authoring workflows and web technologies. JSON-LD, in particular, is effective for embedding structured data in client-side generated content and personalized applications, whereas Microdata and RDFa are often preferred in server-side templates. By offering multiple syntaxes, Schema.org enables broad adoption while preserving interoperability across systems.

Schema.org also follows an incremental approach to complexity, with new elements and refinements introduced gradually across different areas of the schema. Vocabulary extensions are managed through hosted extensions, which are integrated with the core but remain optional, and through external extensions, which are independently maintained by specialized communities. These mechanisms allow domain experts in areas such as news, healthcare, e-commerce, and education to contribute refinements without centralized coordination. Collaboration via GitHub and W3C Community Groups ensures that the vocabulary evolves in response to real world use cases while maintaining coherence and consistency.

### 2.3 Investigative Ontologies

Investigative ontologies provide a structured framework that supports investigative processes by modeling evidentiary objects, actors, actions, and their interrelationships. They formalize provenance, traceability, and contextual attributes across the investigative lifecycle, enabling consistent interpretation of heterogeneous data sources.

#### *Cyber-investigation Analysis Standard Expression (CASE)*

CASE [9] is a community developed standard that offers a structured, ontology-based framework for representing information typically analyzed and shared by humans and systems during digital evidence investigations. The effectiveness of CASE comes from offering a shared language that enables automated correlation and validation of varied information sources, while also ensuring that analysis results can be traced back to their sources, recording when, where and by whom actions were performed on the data.

CASE facilitates cyber-investigations across various domains, including criminal and corporate. The primary objects under analysis are Traces, which are described through associated property Bundles. These property Bundles capture multiple attributes of each Trace, encompassing data sources including mobile devices and storage memory, as well as digital objects such as messages (email or chat), multimedia content, and logs including browser history and event records. CASE builds upon the Unified Cyber Ontology (UCO), which establishes classes of cyber objects (e.g., items, tools, people, places), their relationships with other cyber objects, and the provenance of items and actions performed throughout an action life cycle.

CASE is built to uphold data protection, privacy and secure sharing of electronically stored information. By representing data in a fully structured, ontology-driven format, CASE facilitates sophisticated analytic techniques, including pattern detection, graph-based queries, data mining, abstraction, and visualization, helping investigators pinpoint critical items and reveal hidden connections across cases.

Additionally, CASE provides a comprehensive structure for representing the lifecycle of electronic evidence and the actions performed on it. It captures provenance information, including who performed each action, when it occurred, and the tools or methods used. By linking Traces, Actions, and Identities within its ontology-based framework, CASE ensures that digital evidence can be consistently documented, tracked, and exchanged across systems. This structured representation supports secure sharing, proper handling of sensitive information, and enables more effective analysis by preserving the context and relationships of data throughout an investigation.

***LANTERN – TECH COALITION***

Since perpetrators frequently leverage multiple digital platforms to disseminate abusive content and exploit minors, detection and intervention efforts limited to individual platforms often fall short. Lantern Program addresses this problem by enabling technology companies to securely share threat indicators and intelligence related to child sexual exploitation and abuse, thereby enhancing their ability to identify harm that might otherwise go undetected. In terms of its structure, Lantern should not be construed as a formal ontology. Instead, the Lantern Program consists of a taxonomy, primarily designed for investigative purposes, enabling participating organizations to systematically identify and analyze platform policy violations with accuracy and consistency.

Prior to the Lantern Program, no systematic framework existed to coordinate cross-platform defenses against such predatory actors. Through Meta's ThreatExchange platform, Lantern enables signal sharing among participating platforms [10]. Participants are expected to manually review shared signals and determine violations of their own platform policies before acting. Signals can be shared and accessed via a user interface or an API.

The Lantern Program's efforts are organized around two key components: The Lantern Program Taxonomy and Signals. The Lantern Program Taxonomy functions as a structured classification framework that underpins the platform's signal sharing initiative. It provides a comprehensive set of definitions and categories, including types of CSAM (e.g., animated, generative, manipulated, self-generated) and actors (e.g., coordinators, consumers, distributors), which are implemented through tags to standardize terminology and facilitate consistent interpretation of data. As a living document, the taxonomy is continuously updated to reflect emerging threats, ensure compliance, and maintain the quality and usability of shared information.
By standardizing terminology and categorization, the taxonomy enables participating organizations to systematically analyze and interpret shared data, thereby enhancing the effectiveness of cross-platform investigative efforts. Signals, in contrast, represent concrete instances of potential policy violations that companies upload and share. These can be content-based, such as CSAM hashes, URLs and keywords, or incident-based, such as usernames or email addresses. Each signal is tagged according to the Lantern Program Taxonomy, ensuring consistent categorization and facilitating cross-platform analysis.

While the Lantern Program Taxonomy provides the structured framework for interpreting signals, the signals themselves constitute actionable intelligence that participating organizations can use to identify patterns of abuse, prevent further harm, and coordinate responses, all in accordance with applicable laws and privacy standards. Each signal submitted to Lantern must be tagged with at least one entry from the official Program Taxonomy, which is collaboratively managed by the Tech Coalition and participating members. Developed from multiple inputs and sources, the taxonomy also supports key functions such as compliance, quality assurance, and enhancing the operational value of shared signals.

### 2.4 Comparison

Overall, the reviewed frameworks span three distinct yet interrelated categories, encompassing both formal ontologies and structured taxonomies. CSA/CSE-focused ontologies aim to enhance victim identification and improve the systematic handling of CSAM through harmonized conceptual and classification schemes tailored to the child protection domain.

Metadata oriented ontologies provide structured frameworks for organizing and linking metadata elements that describe the properties and contextual attributes of data, enabling interoperable representations across heterogeneous systems and supporting data reusability. Investigative frameworks support the representation of evidentiary objects, actors, actions, and signals, while ensuring provenance and traceability throughout investigative processes.

The adoption of a particular structure reflects the practical requirements and operational scope of its application. In this respect, each framework aims to reconcile conceptual rigor with operational feasibility, while ensuring secure data handling and adherence to sound information governance principles. Error! Reference source not found. presents a comparative overview of CSA/CSE-oriented, metadata-oriented and investigative ontological frameworks, highlighting their core objectives, functional roles, representative applications and distinct characteristics.

| Framework Type | CSA/CSE-related Ontologies & Classification Systems | Metadata-oriented Ontologies | Investigative Ontologies & Taxonomies |
|---|---|---|---|
| **Primary Objective** | Facilitate victim identification and structured processing of CSAM/CSEM | Organize and describe data resources through standardized metadata, supporting interoperability across systems | Support investigative processes and cross-organizational information sharing |
| **Core Functional Orientation** | Content classification, victim identification | Resource description and semantic annotation | Modeling of evidentiary objects, actors, actions, and investigative workflows |
| **Operational Context** | Hotline analysis, law enforcement casework, and industry reporting | Digital repositories, web platforms, and linked data ecosystems | Digital forensics, cyber-investigations, cross-platform threat coordination |
| **Representative Examples** | Universal Classification Schema (INHOPE)<br>Luxembourg Guidelines<br>Towards a Global Indicator of Unidentified Victims in Child Sexual Exploitation Material (ICSE/INTERPOL)<br>The Tech Coalition's Industry Classification System | Dublin Core – DCMI Metadata Terms<br>Schema.org | Cyber-investigation Analysis Standard Expression (CASE)<br>Lantern Program Taxonomy |

**Table 2**. Comparison of CSA/CSE-oriented, metadata-oriented and investigative ontological frameworks.

The comparative analysis demonstrates that the reviewed frameworks address complementary aspects of the CSA/CSE domain. CSA/CSE-related frameworks primarily focus on content classification and victim identification, metadata-oriented ontologies emphasize semantic annotation and interoperability, whereas investigative ontologies model evidentiary objects, actors, and investigative workflows. Building upon these complementary approaches, the PreventCSA@EU ontology integrates these functionalities within a single semantic framework specifically designed to support the annotation, classification, and management of CSAM/CSEM-related information, while maintaining conceptual alignment with the INHOPE Universal Classification Schema (UCS), Dublin Core DCMI Metadata Terms, and Schema.org.

## 3. Methodology

Building upon the comparative analysis presented in the previous section, the development of the PreventCSA@EU ontology followed a structured ontology engineering process following established ontology engineering methodologies [15] comprising four successive phases: (i) requirements analysis, (ii) ontology conceptualization, (iii) semantic alignment, and (iv) operational integration. This methodology ensured that the resulting ontology addresses the operational needs of national LEAs while maintaining semantic interoperability with widely adopted ontological frameworks.

The development process commenced with the analysis of the functional, operational, and semantic requirements that the ontology should satisfy. These requirements were elicited through a series of stakeholder meetings involving representatives from multiple Directorates of the Hellenic Police responsible for the investigation and management of CSAM/CSEM cases [2]. The meetings focused on identifying existing operational workflows, investigative priorities, annotation practices, information exchange requirements, and the challenges associated with the management of CSA/CSE investigations. Particular attention was also given to the emerging European regulatory framework, including the proposed Child Sexual Abuse Regulation, in order to ensure that the ontology would support future interoperability and interconnectivity with the envisaged EU Centre databases and other relevant European infrastructures.

Based on the identified requirements and the findings of the comparative analysis presented in Section 2, the PreventCSA@EU ontology was designed as a unified semantic framework integrating, among others, concepts derived from CSA/CSE-related classification systems, metadata ontologies, and investigative frameworks. Rather than adopting an existing ontology, the proposed model introduces a hierarchical semantic structure centered around five principal components, which collectively capture the entities and relationships required for the semantic representation, annotation, classification, and management of CSAM/CSEM-related information.

A key design objective of the PreventCSA@EU ontology was to achieve conceptual alignment with the INHOPE Universal Classification Schema (UCS) at the label level. The alignment process was guided by the semantic meaning and definitions of the concepts included in the UCS rather than by establishing structural equivalence between the two models. Wherever appropriate, the PreventCSA@EU ontology adopts the UCS labels to facilitate semantic interoperability and comparability across heterogeneous systems, while its internal hierarchical organization was independently developed to satisfy the operational and investigative requirements identified during the requirements analysis.

This alignment facilitates semantic interoperability and comparability across heterogeneous information systems and jurisdiction specific classification practices through the use of a common classification vocabulary. To further enhance interoperability and metadata compatibility, the ontology also incorporates alignments with Dublin Core DCMI Metadata Terms and Schema.org. This design enables the ontology to satisfy both the operational requirements identified through stakeholder consultations with national Law Enforcement Agencies (LEAs) and the broader interoperability requirements of emerging European infrastructures.

## 4. PreventCSA@EU

### *PreventCSA@EU Conceptual Framework*

The PreventCSA@EU ontology is a formal framework that represents concepts within the domain of CSA and CSE, along with the relationships between them. It provides a structured and standardized approach to improving the identification, classification, and analysis of CSA/CSE-related content and material. Due to the usage restrictions governing access to the Universal Classification Schema (UCS), the PreventCSA@EU ontology is not presented exhaustively at the level of individual classes, properties and scope notes. Instead, the paper focuses on the core conceptual structure of the ontology, its principal semantic relationships and its alignment strategy with external classification frameworks.

In the following section, the conceptual alignment of the PreventCSA@EU ontology with INHOPE's UCS at the label level is highlighted. To support this objective, PreventCSA@EU labels that are conceptually aligned with UCS labels have been incorporated into the core structural elements of the ontology. However, references to the respective UCS aligned denominations are presented descriptively rather than through their exact terminology, due to the access and disclosure restrictions associated with the UCS. Based on the PreventCSA@EU

ontology, the process model for the CSA indicators' database (Annotated Hash Database) is developed to align fully with its conceptual framework, as illustrated in **Figure 1** (Class Hierarchy).

A distinguishing characteristic of the PreventCSA@EU ontology is its coherent hierarchical structure, in which high level abstract concepts are systematically refined into more specialized subclasses. This design enables consistent semantic representation, inheritance of properties, and reasoning across multiple levels of conceptual granularity. In contrast to a taxonomy, which is characterized by lower structural complexity, the PreventCSA@EU ontology exhibits a richer organizational model, supporting both detailed domain analysis and operational database implementation.

The PreventCSA@EU database serves as a practical instrument for implementing the aims of the "Towards a Coordinated and Cooperative Effort for the Prevention of Child Sexual Abuse at a European Level" Action, supporting enhanced cooperation among national authorities, service providers, and EU bodies for the prevention and combating of online CSA, exploitation, and grooming. The design of the Annotated Hash Database based on the ontology aims to enable the effective detection, reporting, and removal of CSAM and CSEM, while ensuring semantic interoperability, operational applicability, and adherence to data protection and responsible information management principles. By defining clear categories, relationships, and hierarchies, the PreventCSA@EU ontology not only supports reasoning about the properties of the CSA/CSE domain, but also serves as a foundation for organizing information in a consistent and systematic manner.

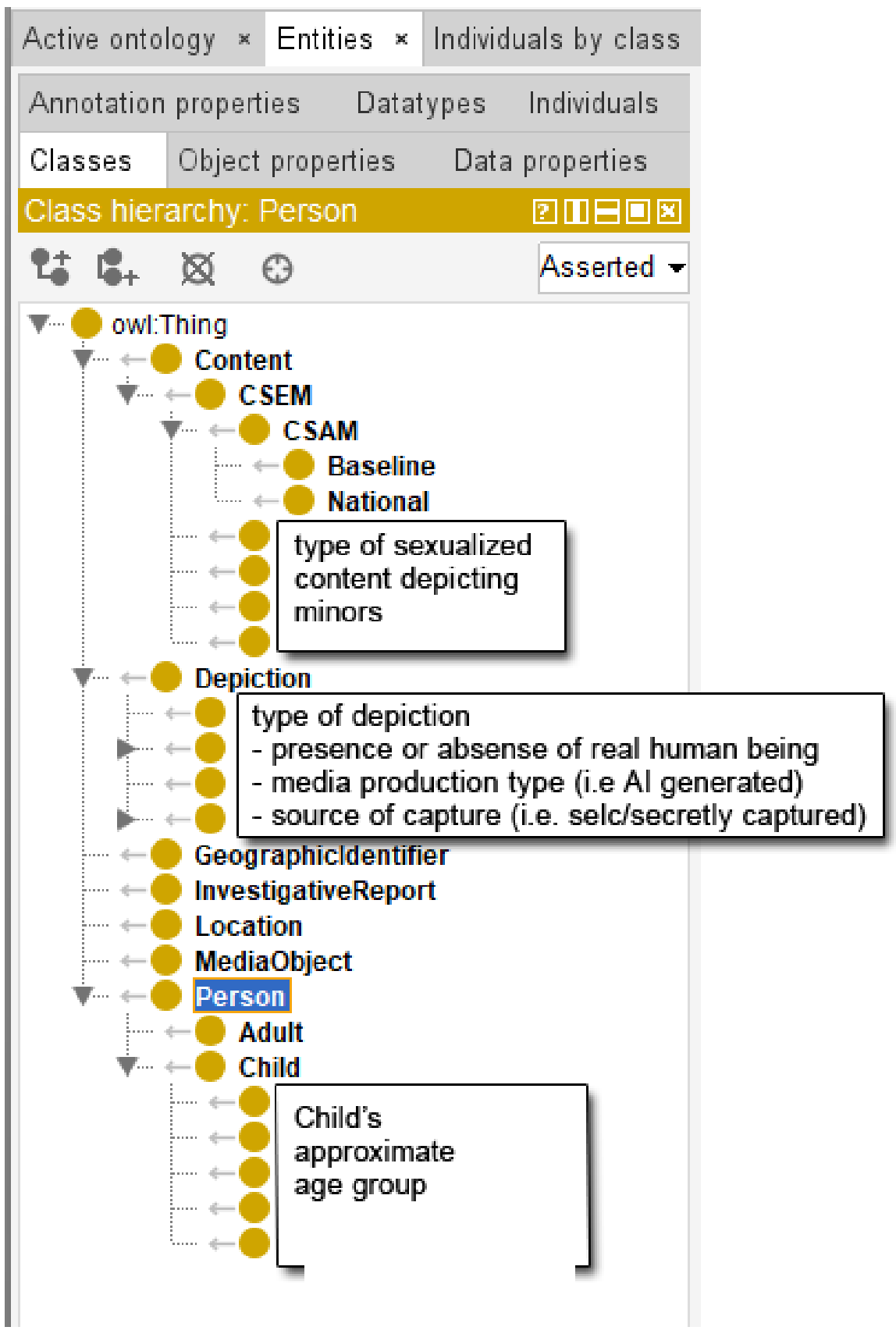

**Figure 1**. Class Hierarchy of PreventCSA@EU Ontology.

The development of this ontology responds to existing challenges in combating CSAM, such as fragmented data, inconsistent terminology, and the absence of shared frameworks. It seeks to establish a common vocabulary for key concepts, terms, and relationships relevant to CSAM/CSEM detection and prevention, enhancing interoperability across platforms, tools, and stakeholders. Additionally, it ensures that technical rigor is balanced with the imperative to handle highly sensitive information ethically and responsibly.

The PreventCSA@EU database will be populated exclusively by the relevant services of the Hellenic Police with hashes and associated annotations, without storing the illegal content itself. The database has adopted the dominant cryptographic (SHA-1 – Secure Hash Algorithm 1, SHA-256, and MD5 – Message Digest Algorithm 5) and perceptual hash algorithms (PDQ and PhotoDNA), in order to ensure compatibility with major hash checking systems such as IWF [11], NCMEC [12], THORN [13] and ICCAM [14].

Hashes are submitted through a web-based client application, which communicates with the Hash Checking Service (HCS) via the Hash Checking API. . The interaction between the HCS and the PreventCSA@EU database will be performed via an API, providing tiered access and supporting the extension and integration of additional trusted data sources, such as the ISPs and the Greek INHOPE Hotline (SafeLine.gr) [16]. The Hash Checking API supports both bulk and individual hash queries in a strictly read only mode, while it does not permit modification or annotation of stored hashes. It solely returns existence/non-existence responses, thereby preserving data integrity and access control constraints.

The overall PreventCSA@EU ecosystem follows a conventional three-tier architecture consisting of a web-based client application, the application layer implemented by the HCS and its API, and the Annotated Hash Database. The client application enables authorized users to submit media objects or hash values individually or in bulk for verification. The PreventCSA@EU database is populated exclusively by the relevant services of the Hellenic Police through authorized ingestion workflows, where newly generated hashes are stored together with their associated semantic annotations.

The API acts as the communication layer between the client application and the AHDB, forwarding requests to the server and returning the corresponding responses. The PreventCSA@EU ontology provides the semantic model underlying the annotation process by defining the structure, concepts, and relationships of the metadata associated with each stored hash. The HCS performs the hash generation and verification operations, querying the AHDB for exact or perceptually similar matches and returning the corresponding verification results. This architecture clearly separates presentation, application, and data management responsibilities while ensuring semantic interoperability, secure access control, and extensibility for the future integration of trusted external data sources.

A representative interoperability scenario involves the exchange of hashes and their associated semantic annotations between the Hellenic Police and trusted Law Enforcement Agencies (LEAs) or partner organizations. Hashes and annotations from authorized external sources may be imported individually or in bulk through CSV files or direct API calls. Before being incorporated into the Annotated Hash Database, all imported data are validated against the PreventCSA@EU ontology to ensure semantic consistency. Conversely, cooperating LEAs and authorized systems may retrieve annotation information through the API using standardized JSON-based data exchange, whereas users with lower authorization privileges are limited to verifying the existence of submitted hashes without accessing the associated annotation metadata.

The development of the ontology begins with the central class Media Object, which is defined based on the Schema.org ontology as follows: a media object, such as an image, video, audio, or text object embedded in a web page or a downloadable dataset, i.e. DataDownload. Note that an Object may have many media objects associated with it on the same web page. For example, a page about a single song (MusicRecording) may have a music video (VideoObject), and a high and low bandwidth audio stream [17]. The usage of the MediaObject class is shown in **Figure 2**.

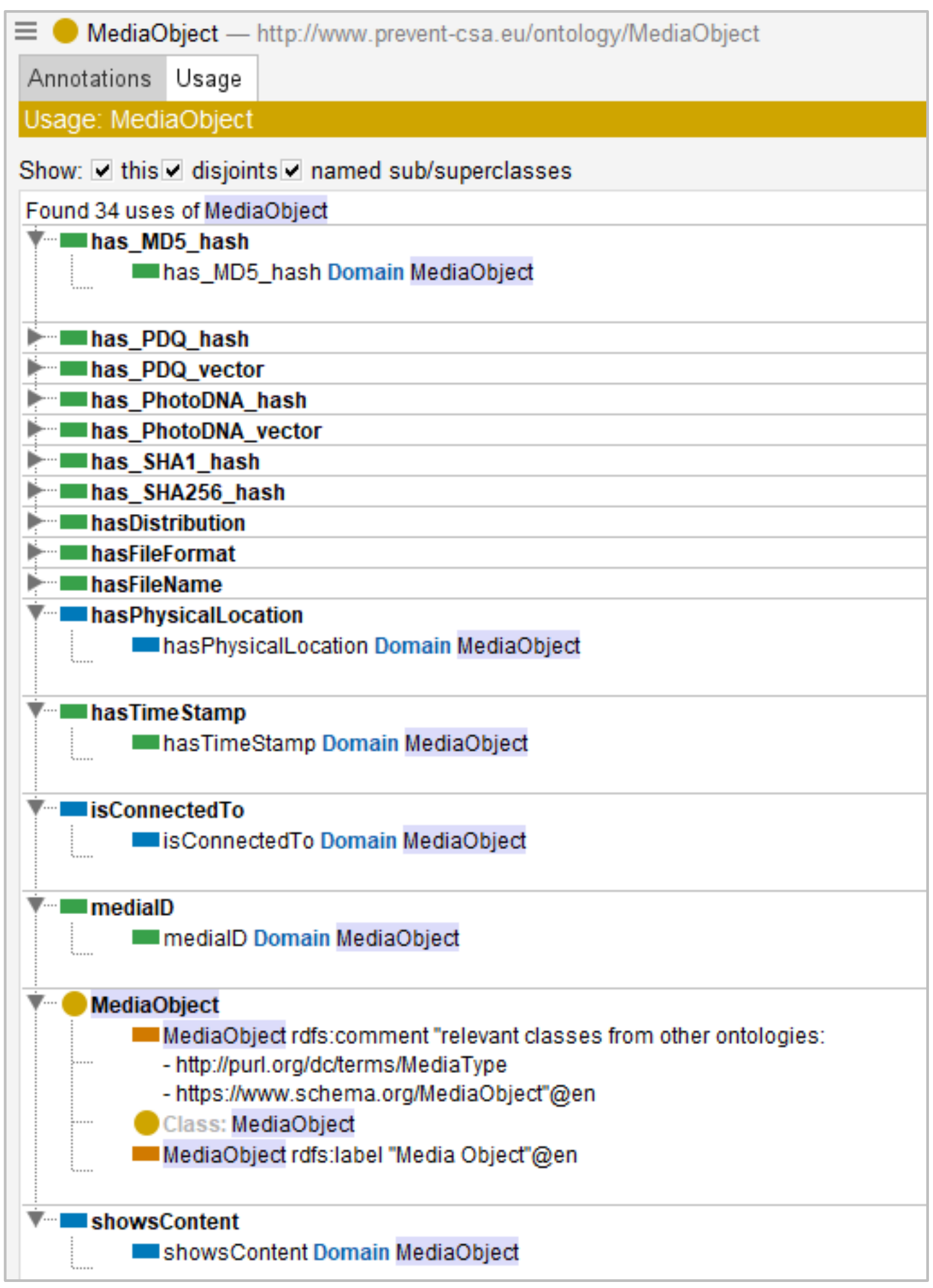

**Figure 2**. Usage of class "Media Object" in the PreventCSA@EU Ontology.

The Media Object class is linked to one or more hash values through a data property (e.g. has_SHA1_hash). A data property of the class is also the File Format, while the Media Object is associated with a Location via the HasPhysicalLocation object property, indicating the server where the file resides. The ontology also incorporates metadata associated with the File Format and the Location of the Media Object. These three classes are primarily derived from two well established ontologies: Dublin Core – DCMI Metadata Terms (DCMI Metadata Terms) and Schema.org. The FileFormat defined as "a digital resource format" is a data property of the Media Object, inspired by the FileFormat [18] from Dublin Core. The formats are specified using the list of Internet Media Types [19].

The distribution characteristics of a Media Object are specified through the hasDistribution data property, which is defined as an enumeration constraining its values to a predefined set of permitted options. This formalization ensures consistency and clarity in representing the distribution characteristics of media entities within the ontology.

Additionally, the Location of the Media Object is defined as "a spatial region or named place". This class aligns with the terminology of the Dublin Core Location class [20] and is analogous to the Place from the Schema.org ontology [21]. The Location class is used to specify the location of the server hosting the Media Object as well as, when available, the geographic location where the content of the Media Object was created or did occur.

Specifically, Location is associated with a Geographic Identifier through the object property HasGeographic Identifier. The Geographic Identifier class encapsulates the precise spatial characteristics of the location via several data properties: coordinate system (text, defaulting to WGS84), coordinate units (text, defaulting to degrees), longitude (decimal), and latitude (decimal). This design allows each Location to reference a standardized geospatial representation, ensuring accurate mapping of spatial information across systems. The application of the location class is illustrated in **Figure 3** below.

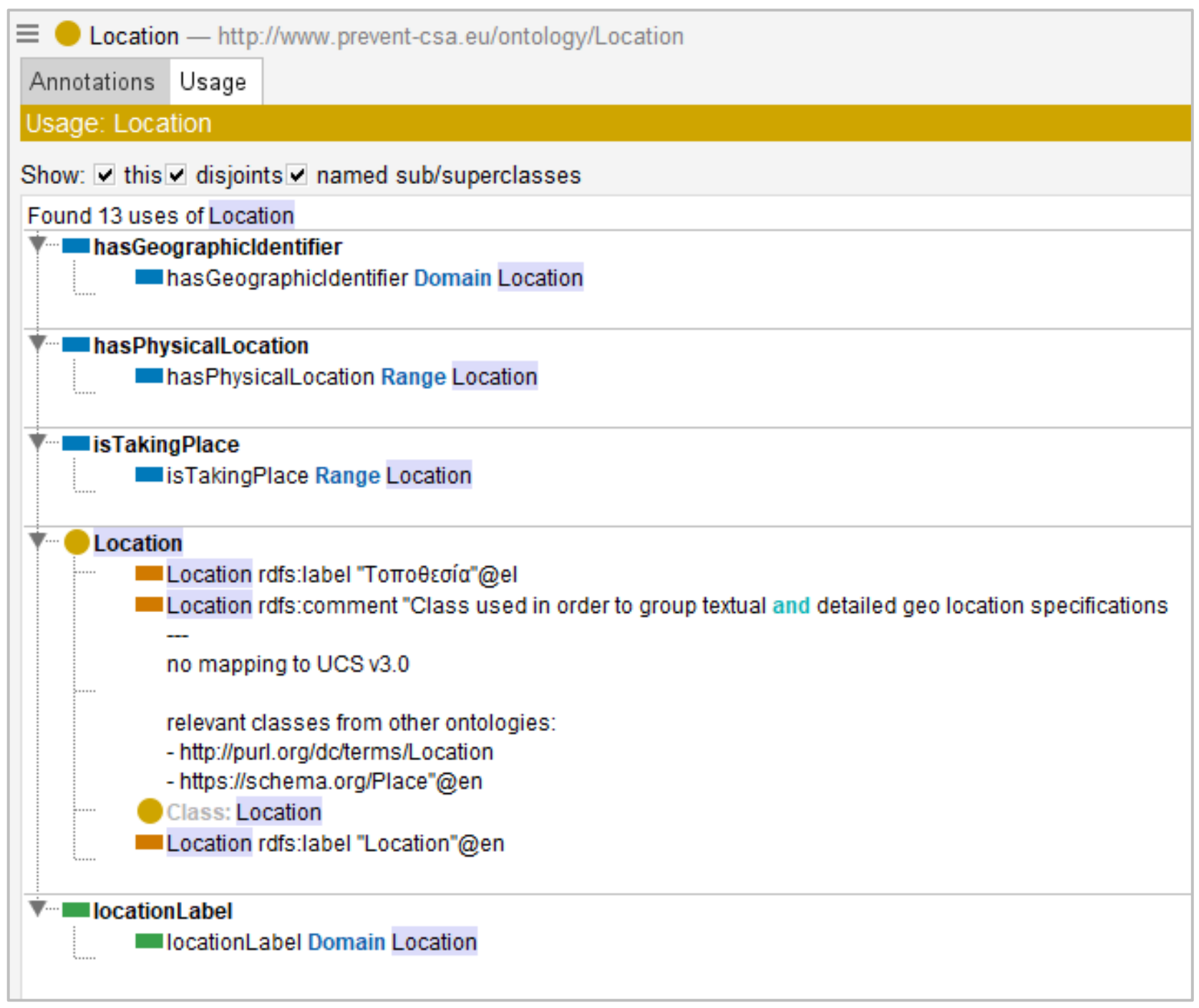

**Figure 3**. Usage of class "Location" in the PreventCSA@EU Ontology.

The second core element of the PreventCSA@EU ontology is Content, which includes either an adult or a child (disjoint Person's subclasses). In the context of our scope, the content of the material relevant to our project will portray Child Sexual Exploitation Material (Class CSEM). The Content may be further classified as CSAM, which is a subclass of CSEM. According to INHOPE, CSAM refers to images or videos depicting a child who is engaged, or represented as being engaged, in explicit sexual activity. In contrast, CSEM constitutes a broader category that encompasses exploitative sexualized content involving minors which may not necessarily be classified as illegal CSAM under national legislation [22].

The third core element of the PreventCSA@EU ontology captures the attributes of Person, where the term Person refers to the individual depicted. Each instance of the class Persons is characterized by Skin Tone and Biological Sex. For the purpose of characterizing Skin Tone, the Fitzpatrick Scale guidelines were utilized as an established reference framework. The class Person within this framework is further categorized into two disjoint subclasses: Adult and Child (see below **Figure 4**). The Child subclass includes data properties that describe specific conditions of the depicted victim, including those related to the degree of nudity depicted and the physical or intellectual disabilities. In particular, the Child data property hasVulnerableState is defined as an enumeration with two possible values: mental incapacity and physical incapacity, thereby distinguishing instances of the class according to the type of vulnerability, whether mental or physical.

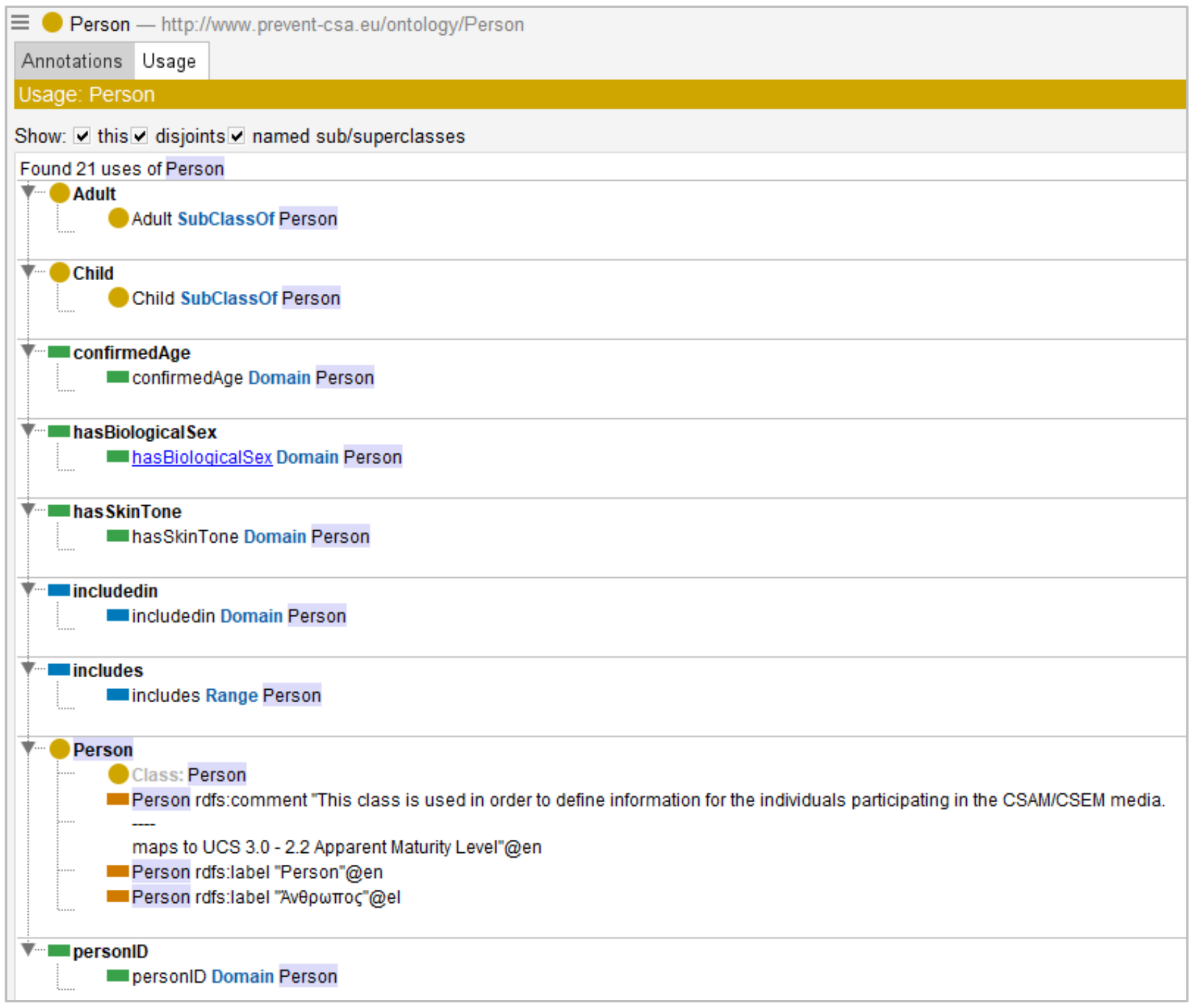

**Figure 4**. Usage of class "Person" in the PreventCSA@EU Ontology.

The fourth core element of the PreventCSA@EU ontology is the Depiction class which represents the depiction of the content in the Media Object. Depiction subclasses are organized around sets of labels, which capture among other aspects, the type of depiction, the presence or absence of a real human being, the mode of media production (including artificially generated content) and the source of its capture, thereby distinguishing between content that was generated by a third party without the knowledge of the depicted subject and content that originated from the depicted subject. Specifically, the labels related to the media's source of capture have been defined in a manner that is broadly consistent with a set of UCS labels, which is used to attribute elements of investigative relevance, from the perspective of law enforcement authorities, to specific CSEM and CSAM material. However, it should be noted that, within the PreventCSA@EU ontology, the respective subclasses are not intended to provide information of direct investigative relevance for LEAs. Consequently, they are not hierarchically incorporated within the 'Investigative Report' element, which is introduced below.

The last element is the Investigative Report, which is connected to the Media Object and comprises a set of data properties that are conceptually aligned with UCS label definitions. These labels are used to record information regarding the offender, the victim, metadata for LEAs, as well as the case file number associated with each media item. The respective annotation values use different datatypes. Some labels are defined as strings, while others are Boolean values representing binary properties.

As already presented, the PreventCSA@EU knowledge base constitutes an independent hierarchical structure, distinct from existing classification schemes, with its primary design aimed at addressing the operational needs and domain specific requirements of national LEA Directorates. Beyond this, its design supports a level of conceptual alignment with UCS label structures at the label level, facilitating cross-organizational understanding, supporting the integration of annotated datasets, and enabling interoperability across contexts ranging from the assessment of illegality to more granular content and context-based use cases, including machine learning and training applications.

This alignment is intended to facilitate comparability across jurisdictions and stakeholder specific classification practices through the use of a shared vocabulary for core classification concepts. While maintaining its independent ontological structure, the design of the PreventCSA@EU ontology also addresses the broader need for a common classification language that can support international cooperation, while simultaneously fulfilling the operational requirements of national LEA Directorates whose representatives were directly consulted during its development.

*Alignment of the PreventCSA@EU Ontology with INHOPE's Universal Classification Schema (UCS)*

As previously emphasized, the PreventCSA@EU ontology (**Figure 5**. The PreventCSA@EU Ontology.below) has been designed to support a degree of conceptual alignment with INHOPE's Universal Classification Schema at the label level. To support this objective, PreventCSA@EU labels that are conceptually aligned with UCS labels have been incorporated into the core structural elements of the ontology. The respective labels that are aligned with the UCS are described below. Access to the UCS was granted by INHOPE for the purposes of supporting the conceptual analysis and design of the PreventCSA@EU ontology. However, the respective denominations are introduced descriptively rather than by their exact names, as they cannot be disclosed due to the usage restrictions associated with the Universal Classification Schema (UCS).

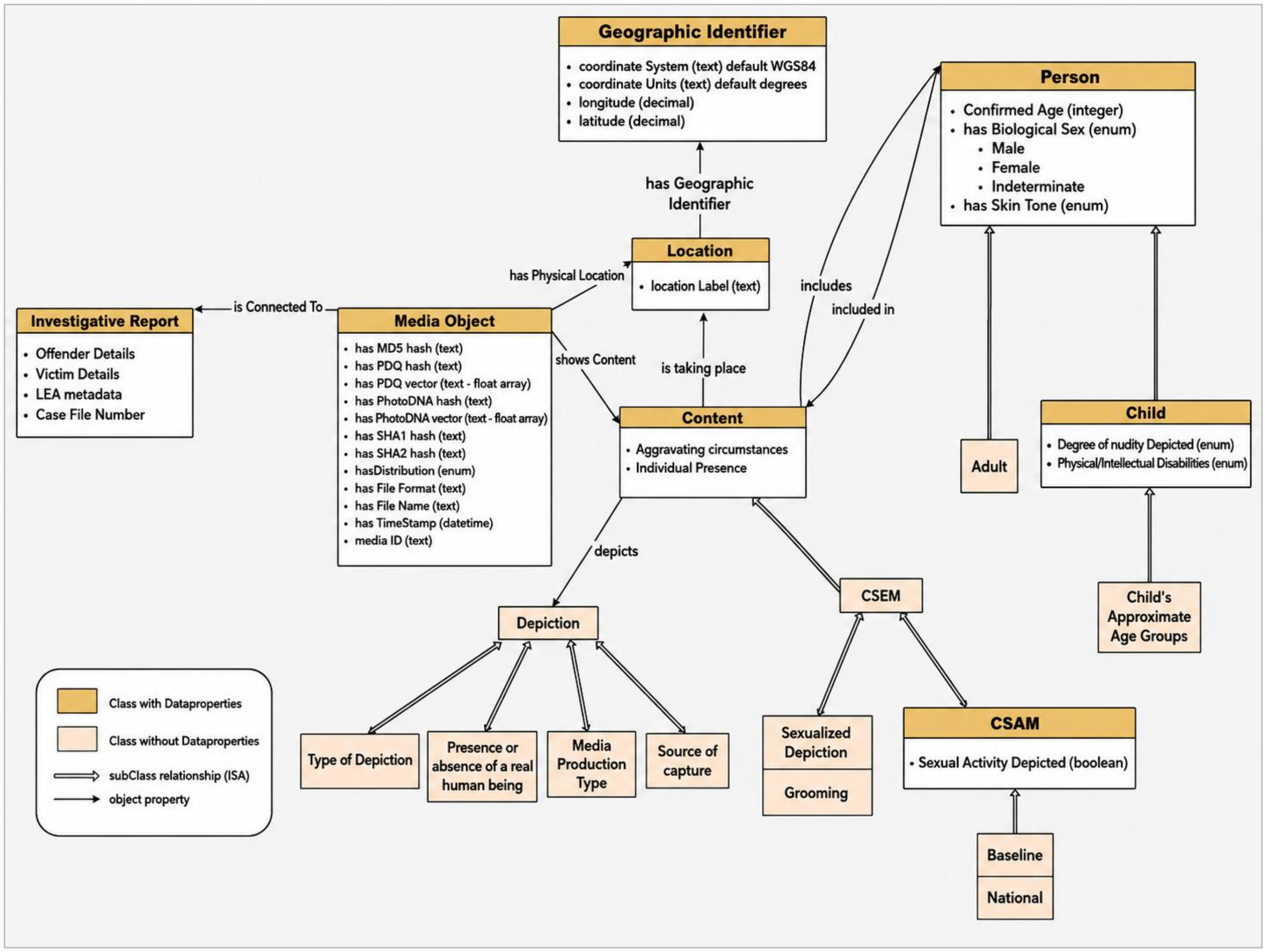

**Figure 5**. The PreventCSA@EU Ontology.

Initially, four subclasses have been introduced under the Depiction class, each of which corresponds to labels associated with labels in the INHOPE UCS. Some of these labels indicate the presence of graphical components added to the media. These components are not considered standalone entities, but rather features that alter or augment the visual representation. Other UCS labels which are incorporated into the PreventCSA@EU ontology indicate the medium used for producing the media.

As previously described, the labels related to the media's source of capture are aligned with a UCS label set used to attribute elements of investigative relevance, from the perspective of law enforcement authorities, to specific CSEM and CSAM material. Additionally, some subclasses with data properties under the Depiction element of the PreventCSA@EU ontology correspond to labels from the Person categorization element of the UCS.

Nevertheless, although these labels have been designed in alignment with the UCS, they are not hierarchically included under either the Investigative Report or the Person elements within the PreventCSA@EU ontology. Instead, they are assigned to the ontology component whose definition best reflects their semantic role within the PreventCSA@EU conceptual model. Consequently, several labels originating from the UCS Person and Investigative elements are modelled as subclasses of the Depiction element, as they describe characteristics of the visual depiction rather than the concepts represented by the Person or Investigative Report components as defined in the PreventCSA@EU ontology.

Under the element Content of the PreventCSA@EU ontology, the subclass CSEM can include subclasses without data properties mapped to the UCS. CSEM also encompasses subclasses that do not originate from the UCS and represent a form of child sexual exploitation. One such example is Grooming. In contrast, CSAM itself, which is a sub-class of CSEM, constitutes a class with data properties mapped to the UCS. These data properties correspond to UCS labels, which are used to indicate the sexual exploitation activities depicted in the media.

Additionally, the Content element of the PreventCSA@EU ontology includes the class Person, which constitutes the third core element of the ontology and is associated with various data properties linked to multiple UCS labels. These labels capture various characteristics of the depicted person. Furthermore, subclasses of Person derived from the UCS indicate whether the depicted individual is a minor, while certain class data properties correspond to UCS labels representing the degree of nudity depicted, as well as potential physical or intellectual disabilities. In addition, specific PreventCSA@EU classes that are conceptually aligned with the UCS categorize the depicted victim into approximate age groups.

The Investigative Report includes various data properties pertaining to investigative information. Some of these properties concern information used to identify the offender, while others relate to information identifying the victim. A further label corresponds to the case file number, which links a given media item to a case under investigation by law enforcement agencies. Some of these data properties correspond to UCS labels, several of which are defined as strings, while others are Boolean values representing binary properties.

Last but not least, the Media Object has a physical Location, linked via a literal to a Class, with data properties linked to the UCS. This class pertains to information that can be used to determine potential geolocation. Although it is aligned with a label class of investigative interest in the UCS, within the hierarchical structure of Pre-ventCSA@EU it is not classified under the Investigative Report. Instead, it is placed under Location, which constitutes an independent class with its own data properties and no hierarchical linkage to the Investigative Report.

## 5. Operational Validation

### *Use Case: Operational Validation of the PreventCSA@EU ontology*

To demonstrate the applicability of the proposed PreventCSA@EU ontology in an operational environment, a representative use case based on the Annotated Hash Database workflow is presented. The objective of this validation scenario is to illustrate how the ontology supports the semantic annotation, structured representation, and consistent classification of previously unrecorded Media Objects while maintaining interoperability with established metadata standards and classification frameworks.

Unlike existing CSA/CSE-related ontologies, which typically focus on either metadata representation, classification, or investigative information, the PreventCSA@EU ontology integrates all these dimensions within a unified semantic framework. The presented use case was selected to demonstrate this capability by illustrating how technical metadata, investigative information, content descriptors, and person characteristics are simultaneously represented through a single ontology-driven annotation workflow.

The operational workflow begins with the submission of a previously unrecorded Media Object to the PreventCSA@EU system. Upon upload, the system automatically generates the available cryptographic (MD5, SHA-1 and SHA-256) and perceptual (PDQ and PhotoDNA) hash values together with the available technical metadata, including filename, file type, timestamp and, where available, geographic information. Before a new record is created, the system verifies that no existing record contains identical cryptographic hash values, thereby preventing duplicate entries while allowing perceptual similarity relationships to be maintained. Following hash generation, the ontology instantiates a new Media Object entity and associates it with the generated digital fingerprints and the corresponding metadata.

During the annotation process, authorized users (Editors and Administrators) are presented with a structured annotation interface organized into three main sections, namely Media–Investigation, Content, and Persons. Although these sections are designed to facilitate the annotation workflow, they directly implement the semantic structure of the PreventCSA@EU ontology. Collectively, they represent the ontology's five principal components, namely Media Object, Content, Person, Depiction, and Investigative Report, exposing them through an operational interface specifically designed to support LEAs annotation workflows.

The Media–Investigation section captures information associated with the Media Object and Investigative Report ontology components. **Figure 6** illustrates the Media-Investigation annotation interface through which authorized users can annotate both automatically generated technical metadata, such as filename, file type, timestamps and available geographic information, and operational information, including the responsible LEA, report identifier, distribution characteristics, investigative status, victim and offender identification indicators, investigative interest, and contextual notes. These annotations instantiate the corresponding ontology classes and properties, providing a semantically consistent representation of both the media object and its associated investigative information.

The Content section corresponds to the Content and Depiction ontology components. Within this section, the submitted media is semantically classified according to the PreventCSA@EU conceptual model by specifying the content category (e.g., CSEM or CSAM), baseline classification, depiction type, aggravating factors, geographic information, and additional semantic descriptors. These annotations instantiate the corresponding ontology concepts while maintaining conceptual alignment with the INHOPE Universal Classification Schema (UCS).

Finally, the Persons section corresponds to the Person component. **Figure 7** illustrates the Persons annotation interface which records structured information for every depicted individual, including person type (adult or child), estimated age, biological sex, and skin tone according to the Fitzpatrick scale. Each depicted individual is instantiated as an independent ontology entity linked through explicitly defined semantic relationships to the corresponding content and media object.

The resulting knowledge graph represents a fully annotated media record in which technical metadata, investigative information, content descriptors, depiction characteristics, and person attributes are integrated into a single semantically consistent representation. During the annotation workflow, all five core ontology components (Media Object, Content, Person, Depiction, and Investigative Report) are instantiated through predefined ontology classes and properties, ensuring that every annotation entered through the platform is semantically represented in a consistent and interoperable manner. The presented use case demonstrates the practical applicability of the proposed ontology by showing how its semantic model supports the complete annotation lifecycle of media records within the PreventCSA@EU Annotated Hash Database, providing a semantic foundation for standardized annotation, interoperability, and ontology-driven CSAM/CSEM data management.

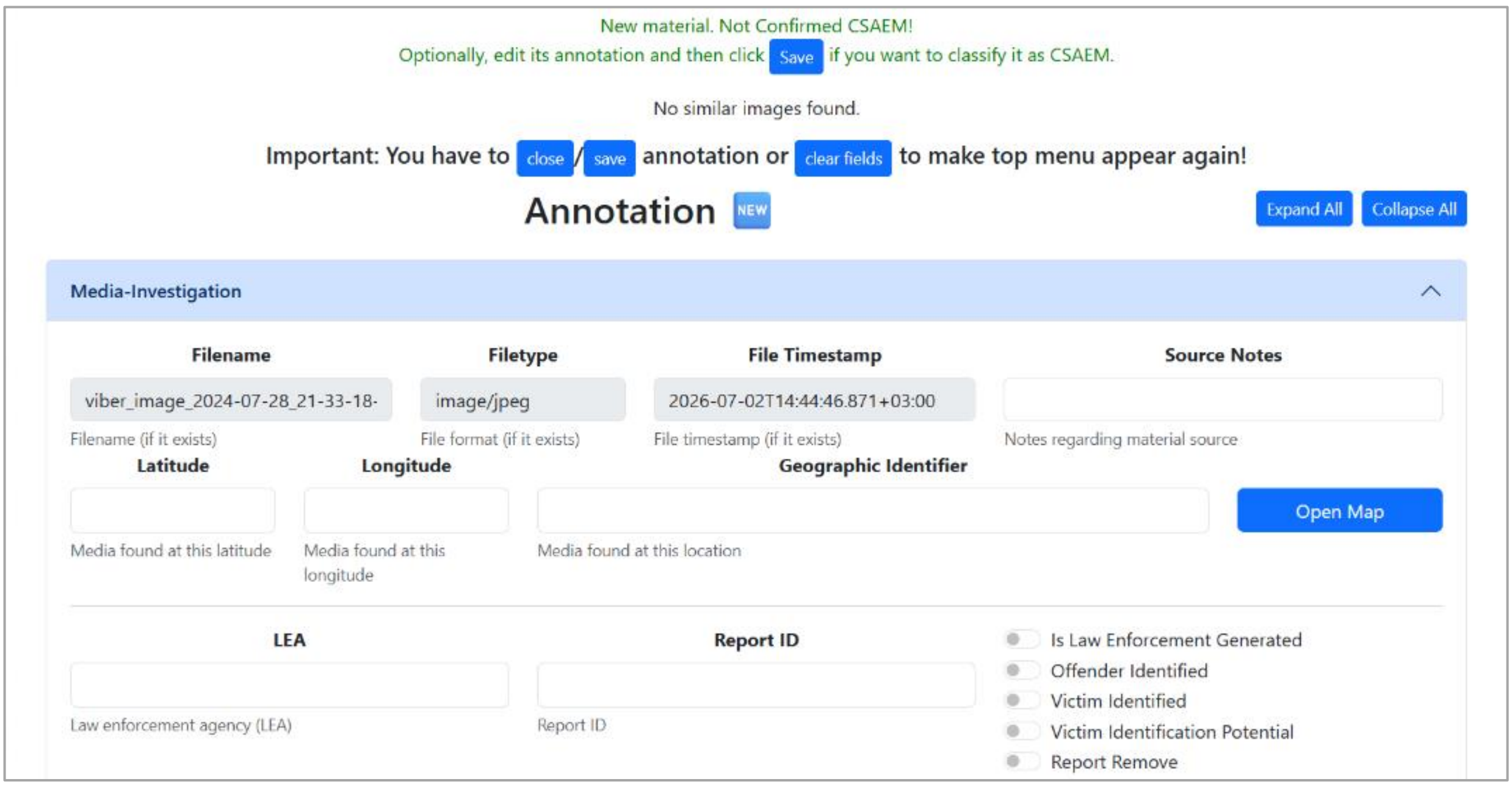

**Figure 6:** Annotation Interface - Media-Investigation Section.

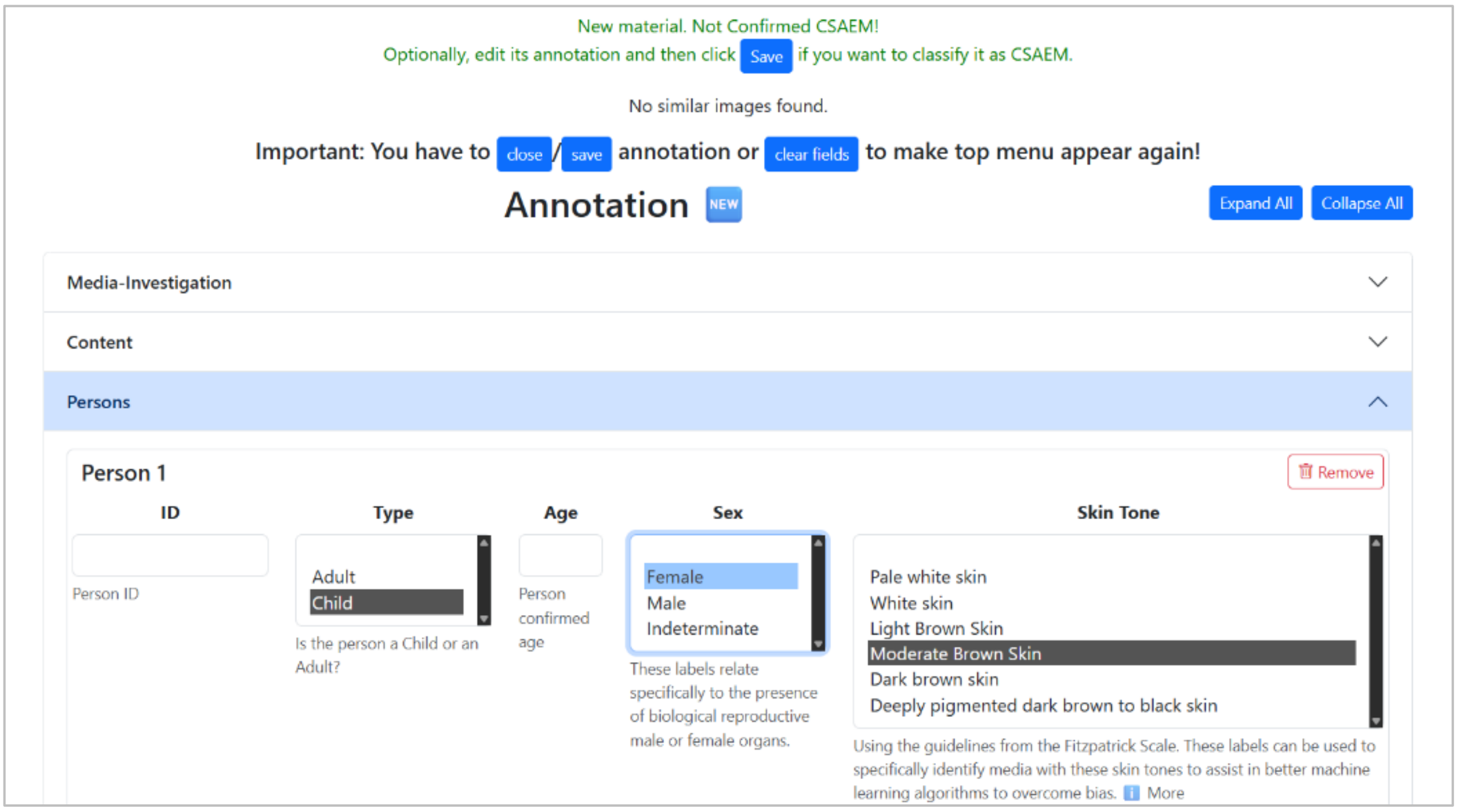

**Figure 7**: Annotation Interface – Person Section.

## 6. Semantic Validation

To semantically validate the proposed PreventCSA@EU ontology framework, a small dataset was extracted from the AHDB and transformed into RDF triples aligned with the PreventCSA@EU ontology. The resulting

RDF data were loaded into a Virtuoso triple store [23], and the ontology was evaluated by assessing its ability to successfully answer the following competency queries.

*Retrieval of content profile statistics*

**Query**: Retrieve aggregated statistics for the *Individual presence* values assigned to instances of `Content` and all its recursive subclasses (the transitive closure of subclasses of Content), while also identifying records with no explicit *Individual presence* annotation.

**Description**: The following query considers all recursive subclasses of Content class and provides statistics on the Individual presence values specified for the Content of each record of the Annotated Hash Database that has been loaded in the virtuoso triple store. The second part of the Union operator identifies records that do not include such a specification value about Individual presence. Similar query can be applied for the statistics retrieval of Content including values for Aggravating circumstances, among other attributes. **Table 3** depicts the results of the query.

```
PREFIX rdfs: <http://www.w3.org/2000/01/rdf-schema#>
PREFIX rdf:  <http://www.w3.org/1999/02/22-rdf-syntax-ns#>
PREFIX pcsa: <http://www.prevent-csa.eu/ontology/>

SELECT
 ?contentClass
 (COALESCE(?interaction, "none") AS ?individualPresence)
 (COUNT(*) AS ?count)
WHERE {
 {
  ?class rdfs:subClassOf* pcsa:Content .
  ?instance a ?class ;
         pcsa:interactionLabels ?interaction .

  BIND(?class AS ?contentClass)
 }
 UNION
 {
  ?media a pcsa:MediaObject .

  FILTER NOT EXISTS {
   ?media pcsa:showsContent ?instance .

   OPTIONAL {
    ?class rdfs:subClassOf* pcsa:Content .
    ?instance a ?class ;
           pcsa:interactionLabels ?interaction .
   }
  }

  BIND("Records without Labels" AS ?contentClass)
  BIND("None" AS ?interaction)
 }
}
GROUP BY ?contentClass ?interaction
```

| contentClass | individualPresence | count |
|---|---|---|
| pcsa:National | Minor / Minor | 1 |
| pcsa:National | Adult present | 1 |
| "Records without Labels" | None | 1 |
| pcsa:Baseline | Minor alone | 1 |

**Table 3**. Content Profile Statistics.

***Retrieval of geographical range results***

**Query**: Retrieve statistics of Media objects found within a region of interest (e.g. 40 km) from the center of Heraklion, Crete, Greece (as specified by coordinates: 25.1344, 35.3403 [24], [25]).

**Description**: The following query retrieves the locations of media objects that are geographically situated within a 40 km radius of a specified reference point. For each location, it returns a human-readable location label (when available), or otherwise the location identifier, the distance from the reference point (expressed in kilometers), and the number of distinct media objects associated with that exact location. The query computes the geographic distance using the latitude and longitude coordinates stored in the ontology and orders the results by increasing distance from the reference location. It can easily be adjusted to return the annotation information of the media objects retrieved within the specified region of interest. **Table 4** depicts the results of this query showing two CSAM/CSEM media objects found within the specified area of interest, whereas **Table 5** lists the results that were excluded due to the range specification of 40 Km from the center of Heraklion.

| locationLabel | distance | count |
|---|---|---|
| “Agia Varvava Municipal Unit, Greece” | 26.56 km | 1 |
| "Kofinas Municipal Unit, Greece" | 33.15 km | 1 |

**Table 4**. Geographical Range results.

| locationLabel | distance | count |
|---|---|---|
| "Τσαγγάρι, Greece" | 600.87 km | 1 |

**Table 5.** Results excluded.

## 7. Conclusions

The PreventCSA@EU ontology provides a structured and semantically precise framework for representing concepts, entities, and relationships relevant to Child Sexual Abuse and Exploitation material. Developed through the structured methodology presented in this work, the proposed ontology combines state-of-the-art classification standards with an entirely novel organizational and hierarchical structure, enhancing extensibility and interoperability while also meeting the operational requirements of national LEAs.

This structured approach allows for the consistent classification, annotation, and analysis of CSA/CSE content, primarily supporting child identification, while also enabling investigative efforts for the prosecution of offenders. The operational validation presented in this work demonstrates the practical applicability of the proposed ontology within the PreventCSA@EU Annotated Hash Database, while the semantic validation further confirms its ability to support meaningful semantic querying and retrieval of annotated information through representative competency queries. Together, these validation activities reinforce the role of the proposed ontology as a critical tool for standardized and semantically consistent CSAM/CSEM annotation, providing a robust semantic foundation for child protection initiatives and the safeguarding of the rights of children and survivors.

A critical aspect of the PreventCSA@EU ontology is its alignment with INHOPE's Universal Classification Schema (UCS), along with Dublin Core – DCMI Metadata Terms and Schema.org. The proposed design enables the derivation of compatibility mappings at the label level, thereby facilitating cross-organizational understanding. At the same time, the ontology incorporates additional independent classes and properties to address operational and investigative requirements that go beyond the schema, thereby balancing standardization with flexibility.

Despite these advances, challenges remain in achieving full interoperability across jurisdictions and stakeholders due to differences in legal definitions and national practices. The adoption of a shared vocabulary is expected to help mitigate these divergences by providing semantic clarity and facilitating the integration of annotated datasets for machine learning, automated detection, and cross-platform analysis.

Although the operational and semantic validation presented in this work demonstrate the practical applicability and semantic consistency of the proposed ontology, further evaluation in large-scale operational environments will strengthen the evidence for its broader impact on semantic interoperability, future machine learning applications, and cross-organizational data exchange. Future work will therefore focus on evaluating interoperability across heterogeneous systems, expanding the semantic validation through additional competency queries and reasoning mechanisms, and investigating the ontology's contribution to ontology-driven annotation and data management workflows.

By leveraging numerous labels from the Universal Classification Schema while extending its structure for operational needs, the PreventCSA@EU ontology establishes a foundation for consistent and interoperable approaches to CSA/CSE content management, promoting both technical rigor and international collaboration in this sensitive domain. At the same time, it supports the interconnectivity of multiple databases operating within the EU, which is particularly desirable in view of the prospective establishment of the EU Centre's databases, should the proposed CSAR Regulation be adopted.

**Author Contributions:**

Conceptualization, E.T., P.F.; methodology, E.T. E.D.; validation, P.F., E.T., E.K. and E.D.; formal analysis, E.T.; investigation, E.T., E.K., E.D.; resources, E.T., E.K., E.D.; data curation, P.F., E.K.; writing—original draft preparation, E.T., E.K., E.D.; writing—review and editing, P.F, E.T., E.K. and E.D.; visualization, P.F., E.K.; supervision, P.F.; project administration, P.F.; funding acquisition, P.F. All authors have read and agreed to the published version of the manuscript.

**Funding**: This research was funded by the Internal Security Fund of the European Union under Specific Action "Towards a Coordinated and Cooperative Effort for the Prevention of Child Sexual Abuse at a European level" (Call: ISF/2022/SA/1.4.1; MIS: 6002910).

**Data Availability Statement**: Restrictions apply to the availability of the Universal Classification Schema (UCS) used in this study. The UCS was obtained from INHOPE with permission and is subject to access restrictions. Therefore, it is not publicly disclosed in this article. Requests for access to the UCS may be submitted through the official INHOPE access request portal: https://forms.office.com/pages/responsepage.aspx?id=iPeAR45LDkygBNMkTuRo5ozQ3Hj9NepBm7TkwpJuM4xUNEJGSElBRk80TU1aSjJVUVBXT1E4TDZIUyQlQCN0PWcu&route=shorturl

**Acknowledgments**: This work was supported by the Internal Security Fund of the European Union under Specific Action "Towards a Coordinated and Cooperative Effort for the Prevention of Child Sexual Abuse at a European level" (Call: ISF/2022/SA/1.4.1; MIS: 6002910). The views expressed are those of the authors and do not necessarily reflect the view of the European Commission.

**Conflicts of Interest**: The authors declare no conflicts of interest.